\documentclass[12pt,twoside]{article}
\usepackage{cite} 
\usepackage{amssymb}

\catcode `\@=11

\newcommand{\eqsnumbering}{ 
\@addtoreset{equation}{section}
 
\def\section{\@startsection {section}{1}{\z@}{-3.5ex plus -1ex minus
     -.2ex}{2.3ex plus .2ex}{\normalsize\bf}}
\def\subsection{\@startsection{subsection}{2}{\z@}{-3.25ex plus -1ex minus
 -.2ex}{1.5ex plus .2ex}{\normalsize\bf}}
 } 
\def\thebibliography#1{\section*{{\normalsize\rm REFERENCES}}
\small\rm\list
 {[\arabic{enumi}]}{\settowidth\labelwidth{[#1]}\leftmargin\labelwidth
 \advance\leftmargin\labelsep\usecounter{enumi}}
 \def\newblock{\hskip .11em plus .33em minus -.07em}
 \sloppy\clubpenalty4000\widowpenalty4000
 \sfcode`\.=1000\relax}

\catcode `\@=12

\newcounter{defthM}
\title{\vspace*{0.5cm} \normalsize\bf  
ON THE DUFFIN-KEMMER-PETIAU FORMULATION OF THE 
 COVARIANT 
HAMILTONIAN DYNAMICS IN FIELD THEORY} 

\author{\vspace{1.5cm} \sc Igor V. Kanatchikov\thanks{E-mail: 
ikanat@ippt.gov.pl, kai@tpi.uni-jena.de}{ }\thanks{On leave from 
Tallinn Technical University, Tallinn, Estonia} \vspace{-1.5cm}\\    
\small Laboratory of Analytical Mechanics and Field Theory \vspace{-0.2cm}\\ 
\small Institute of Fundamental Technological Research\vspace{-0.2cm} \\
\small Polish Academy of Sciences\vspace{-0.2cm}\\
\small \'Swi\c etokrzyska 21,  Warszawa  PL-00-049, Poland \vspace{-0.2cm}\\
\small and \vspace{-0.2cm}\\
\small  Theoretisch-Physikalisches Institut  \vspace{-0.2cm} \\
\small   Friedrich-Schiller-Universit\"at Jena \vspace{-0.2cm} \\
\small   Max-Wien-Platz 1, 07743 Jena, Germany 
}

\date{\small 
\it (%Submitted 
October 1999)   
} 

\begin{document} 

\maketitle 

\markboth{\centerline{\small  I.V. KANATCHIKOV}}{ \hspace*{-12.5pt}
\centerline {\small 
DKP FORMULATION OF COVARIANT HAMILTONIAN DYNAMICS 
%IN FIELD THEORY 
}} 

 \vspace*{-99mm}\vspace*{-2mm} %for 12pt %109 
\hbox to 6.25truein{%\tiny \it    %5.4
\footnotesize\it 
 %%%to appear in   
\hfil \hbox to 0 truecm{\hss 
\normalsize\rm %Preprint  
{\sf  FSU TPI \, 11/99} { }}\vspace*{-3.5mm}}
\hbox to 6.2truein{%\tiny %
\vspace*{-1mm}\footnotesize 
 %%%Rep. Math. Phys. 
\hfil 
} 
%\hbox to 6.25truein{%\tiny 
%\vspace*{0mm}\footnotesize 
 %%%vol. {\bf 4?} (1999)   
%\hfil \hbox to 0 truecm{ 
%\hss \normalsize 
%\sf October 1999 
%\hfil
%}
%} 
\hbox to 6.25truein{%\tiny 
  \footnotesize 
 %%%vol. {\bf 4?} (1999)   
\hfil \hbox to 0 truecm{ 
\hss \normalsize  
\sf hep-th/9911175
}  
}

%\vspace*{61mm} \vspace*{16mm} %for 10pt 
  
\vspace*{67mm} \vspace*{16mm} %for 12pt

 %%%%} 
%%%%%%%%%%%%%%%%%%%%%%%%%%%%%%%%%%%%%%%%%
%% PREPRINTNUMBER %%%%%%%%%%%%%%%%%%%%%%%
%%%%%%%%%%%%%%%%%%%%%%%%%%%%%%%%%%%%%%%%%

\begin{flushright} 
\begin{minipage}{5.5in}
{\small %\footnotesize 
  %\begin{abstract} 
  %\noindent
We show that the De Donder-Weyl (DW) 
covariant Hamiltonian field equations 
 of any field 
can be written in 
 %universal 
Duffin-Kemmer-Petiau (DKP) 
matrix form. 
 %This holds for any field theory which can be represented 
 %as a DW Hamiltonian system. 
 As a consequence, 
 %(generalized)  
 the {(modified)} DKP 
 %$5\times 5$ 
 matrices $\beta^\mu$  
($5\times 5$ in four space-time dimensions) 
are 
   %seen to be 
       of  universal significance for all fields 
 admitting the DW Hamiltonian formulation, 
not only for a scalar field,   
 and can be viewed as field theoretic 
analogues of the symplectic matrix,    
  leading  to 
  %a% %%the %?? 				%%%
  %%``four-symplectic''  			%%% TO LEAVE THIS??? 
  the ``$k$-symplectic'' ($k$=4) structure. 	%%%
We also briefly discuss 
what 
 %would be 
 could be viewed as 
 the %?? 
covariant Poisson brackets 
given by 
 %the ?? 
$\beta$-matrices 
and 
the corresponding 
    %related 
Poisson bracket formulation of DW Hamiltonian  field equations. 
 %\end{abstract}
}   
\end{minipage}
\end{flushright} 
%%%%%%%%%%%%%%%%%%%%%%%%%%%%%%%%%%%%%%%%%%%%%%%%%%%%%%%
%%  END ABSTRACT  %%%%%%%%%%%%%%%%%%%%%%%%%%%%%%%%%%%%%
%%%%%%%%%%%%%%%%%%%%%%%%%%%%%%%%%%%%%%%%%%%%%%%%%%%%%%%

                %%%%%%%%%%%% File ncom.tex %%%%%%%%%%%%%%%
%%%%%%%%%%%%%%%%%%%%%%%%%%%%%%%%%%%%%%%%%%%%%%%%%%%%%%%%%%%%%%%%%%%%%%%%%
 
\newcommand{\beq}{\begin{equation}}
\newcommand{\eeq}{\end{equation}}
\newcommand{\beqa}{\begin{eqnarray}}
\newcommand{\eeqa}{\end{eqnarray}}
\newcommand{\nn}{\nonumber}
 
\newcommand{\half}{\frac{1}{2}}
 
\newcommand{\xt}{\tilde{X}}
 
\newcommand{\uind}[2]{^{#1_1 \, ... \, #1_{#2}} }
\newcommand{\lind}[2]{_{#1_1 \, ... \, #1_{#2}} }
\newcommand{\com}[2]{[#1,#2]_{-}} 
\newcommand{\acom}[2]{[#1,#2]_{+}} 
\newcommand{\compm}[2]{[#1,#2]_{\pm}}
 
\newcommand{\lie}[1]{\pounds_{#1}}
\newcommand{\co}{\circ}
\newcommand{\sgn}[1]{(-1)^{#1}}
\newcommand{\lbr}[2]{ [ \hspace*{-1.5pt} [ #1 , #2 ] \hspace*{-1.5pt} ] }
\newcommand{\lbrpm}[2]{ [ \hspace*{-1.5pt} [ #1 , #2 ] \hspace*{-1.5pt}
 ]_{\pm} }
\newcommand{\lbrp}[2]{ [ \hspace*{-1.5pt} [ #1 , #2 ] \hspace*{-1.5pt} ]_+ }
\newcommand{\lbrm}[2]{ [ \hspace*{-1.5pt} [ #1 , #2 ] \hspace*{-1.5pt} ]_- }
\newcommand{\pbr}[2]{ \{ \hspace*{-2.2pt} [ #1 , #2 ] \hspace*{-2.55pt} \} }
\newcommand{\we}{\wedge}
\newcommand{\dv}{d^V}
\newcommand{\nbrpq}[2]{\nbr{\xxi{#1}{1}}{\xxi{#2}{2}}}
\newcommand{\lieni}[2]{$\pounds$${}_{\stackrel{#1}{X}_{#2}}$  }

\newcommand{\rbox}[2]{\raisebox{#1}{#2}}
\newcommand{\xx}[1]{\raisebox{1pt}{$\stackrel{#1}{X}$}}
\newcommand{\xxi}[2]{\raisebox{1pt}{$\stackrel{#1}{X}$$_{#2}$}}
\newcommand{\ff}[1]{\raisebox{1pt}{$\stackrel{#1}{F}$}}
\newcommand{\dd}[1]{\raisebox{1pt}{$\stackrel{#1}{D}$}}
\newcommand{\nbr}[2]{{\bf[}#1 , #2{\bf ]}}
\newcommand{\der}{\partial}
\newcommand{\oo}{$\Omega$}
\newcommand{\Om}{\Omega}
\newcommand{\om}{\omega}
\newcommand{\eps}{\epsilon}
\newcommand{\si}{\sigma}
\newcommand{\Lm}{\bigwedge^*}
 
\newcommand{\inn}{\hspace*{2pt}\raisebox{-1pt}{\rule{6pt}{.3pt}\hspace*
{0pt}\rule{.3pt}{8pt}\hspace*{3pt}}}
\newcommand{\sro}{Schr\"{o}dinger\ }
\newcommand{\bm}{\boldmath}
\newcommand{\vol}{\omega}%%{\widetilde{vol}}                          
          
         \newcommand{\dvol}[1]{\der_{#1}\inn \vol}
 
\newcommand{\bd}{\mbox{\bf d}}
\newcommand{\bder}{\mbox{\bm $\der$}}
\newcommand{\bI}{\mbox{\bm $I$}}

\newcommand{\ga}{\gamma} 
\newcommand{\Ga}{\Gamma} 
\newcommand{\gmu}{\gamma^\mu}
\newcommand{\gnu}{\gamma^\nu}
\newcommand{\ka}{\kappa}
\newcommand{\hka}{\hbar \kappa}
\newcommand{\al}{\alpha}
\newcommand{\lapl}{\bigtriangleup}

\newcommand{\psib}{\overline{\psi}}
\newcommand{\Psib}{\overline{\Psi}}
\newcommand{\derts}{\stackrel{\leftrightarrow}{\der}}
\newcommand{\what}[1]{\widehat{#1}}
 
\newcommand{\bx}{{\bf x}}
\newcommand{\bk}{{\bf k}}
\newcommand{\bq}{{\bf q}}
 
\newcommand{\omk}{\omega_{\bf k}}
 
%%%%%%%%%%%%%%%%%%%%%%%%%%%%%%%%%%%%%%%%%%%%%%%%%%%%%%%%%%%%%%%%%%%%%%%%%
%%%%%%%%%%%%%%%%%%%%%%%%%%%%%%%%%%%%%%%%%%%%%%%%%%%%%%%%%%%%%%%%%%%%%%%%%
%%%%%%%%%%%%%%%%%%%%%%%%%%%%%%%%%%%%%%%%%%%%%%%%%%%%%%%%%%%%%%%%%%%%%%%%%

\medskip 

In the end of the thirties is was observed \cite{dkp}
 %by Petiau \cite{}, Kemmer \cite{}, 
 %Duffin \cite{} and Japanese \cite{} 
 %... 
 (see \cite{dkp-hist} for  historical details)  
that the first order form of the Klein-Gordon and the Proca 
field equations can be represented in the Dirac-like  
matrix form 
%(for  review see 
%\cite{pauli,corson,umezawa,roman,..}, 
\beq
\beta^\mu \der_\mu U - m U = 0,   
\eeq 
 %The corresponding 
where $\beta$-matrices  
 %(=Duffin-Kemmer-Petiau (DKP) matrices) 
 %were found to 
fulfill the relation 
\beq
\beta^\mu \beta^\nu\beta^\lambda 
+ \beta^\lambda\beta^\nu\beta^\mu 
= \beta^\mu \delta^{\nu\lambda }  
+ \beta^\lambda \delta^{\nu\mu}      
\eeq  
 %where 
 %(with $\eta^{\mu\nu}$    
 %denoting the Minkowski space-time metric)       
which defines the 
so-called Duffin-Kemmer-Petiau (DKP) 
 %DKP 
algebra 
(for a    textbook introduction 
 %can be found, 
 see, e.g., 
\cite{texts,text2}).  

Later on similar considerations 
 %were 
have been 
extended to 
 %the %? 
 %massless fields \cite{} 
higher-spin theories \cite{bhabha,fujiwara,corson},  
gauge fields \cite{bogush,gauge},  
 curved space-time \cite{curved}, 
and even to the Einstein gravity \cite{bogush,fedorov-grav}  
 %supersymmetry 
and arbitrary non-linear equations 
(see \cite{text2,bogush,fedorov-grav} and the references  therein).   
%%%Generalizations to curved space-time were considered in 
%%%\cite{curved}.   
{}From a more mathematical point of view DKP 
 algebras have been considered, e.g., in \cite{corson,chandra,math1,math2}.   
%Mathematical discussion of %Duffin-Kemmer 
%DKP algebras can be found, 
 %have been  presented, 
 %is given 
%e.g., in \cite{corson,chandra,math1,math2}.   

 The starting point of DKP formulation (1) 
is 
 %essentially  
the 
Lagrangian framework since 
 the components of the wave function 
$U$ in (1) are 
%for the scalar field) 
 essentially 
the field variables 
  %$\phi$         
and their first space-time derivatives. 
  %$\phi_\mu$. 
For example, for the scalar field $\phi$ 
eq. (1) is 
a 
matrix formulation of 
 %the system of equations    
the first order form of the Klein-Gordon equation 
$\square \phi = m^2 \phi$: 
\beqa    
\der_\mu \phi &=& m \phi_\mu , \nn \\         
\der_\mu \phi^\mu &=& m \phi         
\eeqa     
%which is the first order form of the Klein-Gordon equation 
  %$\eta^{\mu\nu} \der_\mu \der_\nu y = m^2 y$.   
%$\square \phi = m^2 \phi$.
A particular representation 
of $\beta$-matrices arises from 
rewriting (3) in the form (1) 
using the  
 %five-dimensional 
column wave function 
$U:=(\frac{1}{m}y_\mu,y)^T$.

Recently, we have argued that another first order form of 
the field equations is of interest in field theory 
 %for,  
since  
it represents a manifestly covariant generalization of 
the Hamiltonian formulation from mechanics to field 
theory \cite{ikanat}.  
We mean the so-called De Donder-Weyl (DW) formulation 
\cite{dw} which 
 %arises 
can be summarized as follows. 
 %For 
Let us consider a field theory given by the Lagrangian 
density $L=L(y^a,y^a_\nu, x^\mu)$ which is  a function 
of field variables $y^a$, first derivatives (jets) of fields 
$y^a_\nu = \der_\nu y^a$,   
and space-time variables $x^\mu$ $(\mu=1,...,n)$. Then   
we can define a set of variables 
$p_a^\mu:=\der L/\der y^a_\mu$, 
called {\em polymomenta}, and the 
analogue of the Hamilton canonical function 
$H:=p_a^\mu y^a_\mu -L=H(y^a, p^\mu_a,x^\nu)$, 
called the 
{\em DW Hamiltonian function}, 
such that the Euler-Lagrange field equations 
take the { DW Hamiltonian form} 
\beqa
\der_\mu y^a &=& \der H / \der p_a^\mu ,  \nn \\
\der_\mu p^\mu_a &=& - \der H/ \der y^a .  
\eeqa 
A comparison of (4) with (3) 
suggests 
an idea  
%a possibility 
 %to represent 
 of representing  
the DW Hamilton\-ian equations (4) 
in a matrix form similar to (1).  
 
 %Before doing this 
Let us first 
 %remind  
discuss 
what would be an analogue of 
the procedure under discussion 
for a mechanical system  with a single degree 
of freedom. In this case we have 
 the  configuration space variable $q$ 
and the conjugate momentum 
variable $p$ 
which fulfill the  Hamilton equations of motion 
\beqa 
\der_t q &=& \der H / \der p ,  \nn \\ 
 %\quad  %\nn  \\ 
\der_t p &=& - \der H / \der q  . 
\eeqa
If we introduce a column matrix corresponding 
to the phase space 
vector $z:=(p,q)^T$  equations (5) 
can be written in the matrix form 
\beq
\der_t z = \omega^{-1} \der_z H , 
\eeq
where 
$$\omega := \left( \begin{array}{rcl} 
0&&1 \\
-1&& 0 \\
\end{array}
\right)
$$ is the symplectic matrix. 
Alternatively, one can rewrite 
(6) in the form resembling (1) 
\beq
\omega^{}\der_t z = \der_z H . 
\eeq

The objective of this paper is to 
 demonstrate 
 %argue 
 %show 
that  the DW Hamiltonian 
field equations (4) can be written in the form of DKP equations 
and 
 to argue 
that in this case $\beta$-matrices can be viewed as field theoretic 
generalizations of the symplectic matrix in mechanics.

%From now on 
Henceforth we assume for simplicity  
 %that 
the space-time to be 
four-dimen\-si\-onal 
 %Min\-kow\-ski, ${\mathbb R}^{3,1}$  
 Euclidean, ${\mathbb R}^{4}$   
(a generalization 
to other dimensions and signatures is rather straightforward). 
First, we  consider a 
 %real 
scalar 
field 
theory with a single field variable $y \in {\mathbb R}$. 
Let us introduce the 
five-component 
quantity 
$Z^v:= (p^1,p^2,p^3,p^4,y)^T$${}\in {\mathbb R}^{4} \oplus  {\mathbb R}$, 
$v=1,2,...,5$ (for clarity, we drop the dimensionful factors 
which may be easily restored when necessary).  
Then, similarly to (3) and (1), 
the DW Hamiltonian equations (4) can be 
 %cast to 
 represented in 
the matrix form 
\beq
\beta^\mu_{uv} \der_\mu Z^v = \der H / \der Z^u , 
\eeq
where $\beta$-matrices (= ``the modified DKP matrices'') 
  %(= ``pseudo-DKP matrices'')  
are as follows 

%\medskip 

\begin{center}
$\beta^1=$ $\left(
\begin{array}{ccccc}
0&0&0&0&1 \\
0&0&0&0&0\\
0&0&0&0&0\\
0&0&0&0&0\\
-1&0&0&0&0
\end{array}
\right)$ 
 $\quad \beta^2=$ $\left(
\begin{array}{ccccc}
0&0&0&0&0 \\
0&0&0&0&1\\
0&0&0&0&0\\
0&0&0&0&0\\
0&-1&0&0&0
\end{array}
\right)$  
\end{center}
\beq 
{}\vspace*{-10pt} 
\eeq
\begin{center}
$\beta^3=$ $\left(
\begin{array}{ccccc}
0&0&0&0&0 \\
0&0&0&0&0\\
0&0&0&0&1\\
0&0&0&0&0\\
0&0&-1&0&0
\end{array}
\right)$   
$\quad \beta^4=$ $\left(
\begin{array}{ccccc}
0&0&0&0&0 \\
0&0&0&0&0\\
0&0&0&0&0\\
0&0&0&0&1\\
0&0&0&-1&0
\end{array}
\right)$   . 
\end{center}

%\medskip 

{}From the  comparison of (8) with (7) it becomes obvious that 
$\beta$-matrices (9) are in a sense field theoretic 
analogues of the 
 %(inverse of the) 
symplectic matrix 
in mechanics.  More specifically, they represent the 
 so-called ``$k$-symplectic'' 
 %(here  $k=4$) 
structure given by the 
set of $k=4$ symplectic forms 
$\Omega^\mu:=dp^\mu \wedge dy$:\footnote{Note, that 
the 
 %%a ?? 
 %related 
notion of the 
``$k$-symplectic'' structure 
 %(here  $k=4$) 
was introduced in \cite{awane} 
(see also \cite{deleon} and the references therein). 
Recently it found interesting applications in  studying 
 %pde-s 
  wave propagation in open systems \cite{bridges}. 
Note also that our $\beta$-matrices are 4-dimensional 
analogues of matrices {\bf K} and {\bf M} 
introduced in \cite{bridges} 
in two space-time  dimensions,  
  %with 
our $H$ corresponding to $-S$ in \cite{bridges}. }    
\beq
%dp^\mu_a \wedge dy^a = 
\Omega^\mu = 
\beta^\mu_{uv} dZ^u\wedge dZ^v . 
\eeq 
 %%CHECK THE SIGN HERE 

By direct evaluation one can establish 
that $\beta$-matrices fulfill 
the following relation  
\beq
\beta^\mu \beta^\nu\beta^\lambda 
+ \beta^\lambda\beta^\nu\beta^\mu 
= -  (\beta^\mu \delta^{\nu\lambda }  
+ \beta^\lambda \delta^{\nu\mu})  
\eeq  
which is similar to the defining relation of the DKP algebra, eq. (2),  
up to a sign factor in the right hand side. 
 %%MODIFIED 8.11.99 %%%%%%%%%% 
\newcommand{\explanation}{   
Note that in (11) 
$
(\beta^\mu \beta^\nu\beta^\lambda)_{vv'''} := 
\beta^\mu_{vv'} (\beta^\nu){}^{v'v''}\beta^\lambda_{v''v'''} 
$ 
and the indices $v$ are manipulated by means of 
the 
 %space-time 
metric on ${\mathbb R}^{3,1}$: $diag(+++-)$,  
 %and the field space metric on ${\mathbb R}$,  
 so that, e.g., 
$Z_v=(p_1, p_2, p_3, -p_4, y)$ and 
$\beta^i_{vv'}=+(\beta^i){}^{vv'}$ 
for $i\neq 4$  
and $\beta^4_{vv'}=-(\beta^4){}^{vv'}$.  
 } 
%%%%%%%%%%%%%%%%%%%%%%%%%%%%%%%%%%%%%%%%%%%%%%%%%%%%%%%
%%%%%%%%%%%%%%%%%%%%%%%%%%%%%%%%%%%%%%%%%%%%%%%%%%%%%%%
%%%%%%%%%%%%%%%%%%%%%%%%%%%%%%%%%%%%%%%%%%%%%%%%%%%%%%% 
The minus sign in (11) could be 
absorbed in redefinition of $\beta$-matrices: 
$\beta \rightarrow i \beta$, 
 but in the present context this redefinition seems to be unnatural and 
%but this 
 will not be used in what follows.

%%%%%%%%%%%%%%%%%%%%%%%%%%%%%%%%%%%%%%%%%%%%%%%%%%%%%%%
\newcommand{\replacedtext}{ 
The latter could be 
absorbed by 
 the %? 
redefinition of $\beta$-matrices: 
$\beta \rightarrow i \beta$ which we, however, will not use 
in what follows.  
Note also that our $\beta$-matrices are anti-Hermitean, 
as opposite to the DKP matrices which are 
 known to be 
Hermitean.  }  
%%%%%%%%%%%%%%%%%%%%%%%%%%%%%%%%%%%%%%%%%%%%%%%%%%%%%%%

As a consequence of (11) we obtain the properties of 
 %pseudo-DKP 
$\beta$-matrices 
\beqa 
\beta^3_\mu &=& - 
   %\eta_{\mu\mu}
    \beta_\mu  , \nn \\
\beta_\mu \beta^2_\nu + \beta^2_\nu\beta_\mu 
&=& - 
    %\eta_{\nu\nu}
     \beta_\mu  \quad (\mu\neq\nu) ,  \nn \\
\beta_\mu \beta_\nu \beta_\mu &=& 0 \quad (\mu\neq\nu)  , \nn \\
\beta^2_\mu \beta^2_\nu &=&  \beta^2_\nu \beta^2_\mu  
\eeqa
 %These properties 
which are 
 %also 
similar, up to a sign,  
to 
 %the corresponding 
%the well-known properties of 
%Duffin-Kemmer 
 %those 
 the usual properties of 
 %the usual 
 DKP matrices (see, e.g., \cite{dkp,texts}).

 %%????????? INCLUDE OR NOT ????????? 
\newcommand{\notinclude}{
Note that, alternatively, DW Hamiltonian equations can be 
written in the form  
\beq
\beta^\mu \beta^\nu \der_\nu Z = 
- \beta^\mu \frac{\der H}{\der Z}
\eeq
} %%%

The extension of the present formulation to 
 the  %?  
multicomponent fields 
 %will be 
is different from the usual way the DKP formulation is extended 
from scalar to vector fields (cf. \cite{texts}). 
 Instead of a multicomponent column variable here 
we introduce 
the %?  
  $5 \times m$ matrix, $Z^{v}{}_a$,   
of field variables $y^a$ $(a=1,...,m)$ and 
%their first derivatives $y^a_\mu$ 
 %the corresponding 
polymomenta $p_a^\mu$ 
\beq
 Z^{v}{}_a := 
\left(
\begin{array}{ccccc}
p_1^1&p_2^1&...&p_m^1\\
p_1^2&p_2^2&...&p_m^2\\
%.&.&...&.\\
%.&.&...&.\\
%.&.&...&.\\
p_1^3&p_2^3&...&p_m^3\\ 
p_1^4&p_2^4&...&p_m^4\\ 
y_1&y_2&...&y_m \\
\end{array}
\right)    . 
\eeq 
Then 
%it is easy to see that 
the DW canonical equations (4)  
for an arbitrary $m$-component field 
can be written in the form 
\beq
\beta^\mu_{uv} \der_\mu Z^{v}{}_a = \der H / \der Z^{ua} . 
\eeq 
 %%LOWER/UPPER INDICES ??? 
We conclude, therefore,   
that 
 our $5\times 5$ 
$\beta$-matrices 
have a universal significance for all fields 
which can be 
 represented  in DW canonical form (4), 
not only for a one-component scalar field as 
 %usually   
%is implied by 
   %in the case of 
   with %%! 
the usual DKP formulation.  

A similarity  between $\beta$-matrices and the symplectic matrix 
in mechanics allows us to introduce the 
 %covariant Poisson bracket operation 
 bracket operation 
 of functions of variables 
$Z^u$:  
\beq
\{F,G  \}^\mu := \frac{\der F}{\der Z^u}
(\beta^\mu)^{uv}\frac{\der G}{\der Z^v}  , 
\eeq
which is antisymmetric,  
 %and  
fulfills the Leibniz rule and 
a generalization of the Jacobi identity: 
\beq
\{ \{F,G \}^{(\mu} K\}^{\nu )} + cyclic (F,G,K) = 0 ,   
\eeq 
where $A^{(\mu}{}^{\nu)}:= \half(A^{\mu}{}^{\nu} + A^{\nu}{}^{\mu})$, 
 and thus could be considered as an analogue of the Poisson bracket. 
 %%(cf. \cite{good,tapia}). %%ALTERNATIVE TO SAVE ONE LINE 
  %This bracket, 
When written in 
 %the %? 
explicit form  this bracket  coincides with  the one put forward earlier 
by Good \cite{good} and Tapia \cite{tapia}.  
  %However, 
However, the potential 
 %difficulty 
 problem 
 with the bracket (15) 
is that it maps 
two functions to 
  %, essentially,  
 a vector, 
and that it is not 
 %clear 
 obvious what is the mathematical meaning of the analogue of 
the Jacobi identity, eq. (16).   
 Besides,   although 
 the %? 
bracket (15) allows us to write 
the DW Hamiltonian equations (4) 
in  
  %Poisson   
bracket formulation:  as a consequence of (8) and (15) we obtain 
\beq
\beta^\mu \beta^\nu \der_\nu Z = \beta^\mu \der H / \der Z = 
\{ Z, H \}^\mu,  
\eeq 
  it is not clear 
 %whether or not 
 %this Poisson bracket formulation 
how  this  formulation could be 
extended to more general observables (here,  functions of $y^a$ and 
$p_a^\mu$) 
 to be a proper generalization of 
$\der_t F = \{H, F \}$ in mechanics. 
% is an open question.  
 
Alternatively, one could  
try to 
introduce a bracket operation 
which maps two vectors to a vector 
(cf., e.g., the last equation in \cite{sardan}). 
 %by properly contracting two of three vector 
 %indices appearing in the bracket 
   %therewith 
 %in this case 
 %the last equation in \cite{sardan}). 
 However, 
 %the specific 
  a satisfactory %%?? 
formula consistent with a version of 
the Leibniz rule  and the Jacobi idenity 
 %properties 
so far is not known to us.  
 %(cf., e.g., the last equation in \cite{sardan}). 
%%%%%%%%%%%%%%%%%%%%%%%%%%%%%%%%%%%%%%%%%%%%%%%%%%%%%%% 
\newcommand{\bracket}{
However, a satisfactory 
 %realization of this possibility 
 implementation of this idea  
 %is still to be found. 
 so far is not known to us.  %
%However,  no satisfactory realization of this possibility 
%seem to be  available so far.  
  %the author has no conclusive results in this direction.  
}
%%%%%%%%%%%%%%%%%%%%%%%%%%%%%%%%%%%%%%%%%%%%%%%%%%%%%%%%%

Let us recall that 
in our previous papers \cite{ikanat} 
 another Poisson bracket operation 
for the De Donder-Weyl formulation %theory 
 %(defined on differential forms) 
was introduced 
 %from a %different 
 %based on 
 using a more geometric point of view. 
 The bracket  
is defined on differential forms representing 
observables and 
 is 
based on the notion of 
 what we call 
the polysymplectic form. 
Whether or not a connection 
with the constructions of the present paper 
can be 
 %found 
 established 
is 
 %an open 
 %still 
a question yet to be investigated. 
In fact, a relationship between DKP and Clifford algebras  
\cite{math1,math2}  
and 
a  %the? 
 geometric character of the latter, 
%could 
 %indicate to 
%imply such a connection.
 %%The latter 
%It is also suggested by 
 as well as a relation between the 
 ``$k$-symplectic''  %($k$=4) 
 %four-symplectic 
structure (10) and the polysymplectic form 
used in \cite{ikanat},  
imply such a connection.  

%Obviously, 
 %another 
 A natural question is whether or not the 
 structures 
 %constructions 
described in the present paper, 
 %%when %? 
viewed as field theoretic 
generalizations 
of the 
 %corresponding 
 structures 
 %constructions 
of the Hamilton\-ian formalism in mechanics,  
could be useful  
for an inherently covariant canonical quantization in field theory.  
 %For example, 
 %The starting 
A natural starting point of 
 %such a quantization 
quantization of this type  
%could 
would be the 
``canonical bracket''  following from (15) 
\beq
\{p_a^\mu, y^b \}^{\nu} = 
- \delta^b_a \delta^{\mu\nu} . 
\eeq 
However, at  present  it is not known 
 %is not clear 
  how the Dirac quantization rule 
  should be generalized   
to a $\nu$-indexed bracket.  
  %For the sake of completeness 
 Let us note that 
  %another, conceptually related, 
  a conceptually related 
approach  to field quantization based on 
DW Hamiltonian formulation  can be developed 
\cite{ikanat2} 
using the 
afore mentioned (graded)  
Poisson bracket on differential forms.  
 %\cite{ikanat2}. 
% has been  discussed in 
%\cite{ikanat2}.  

%\medskip 
\bigskip 

{\bf Acknowledgments:}   
{\footnotesize 
I thank V.M. Red'kov for illuminating 
 %remarks 
correspondence and 
 %critical remarks,    
 comments, 
and for drawing 
my attention to the work on DKP formulation 
done in Minsk (Belarus) in the sixties-seventies 
(reviewed in \cite{text2,bogush}).   
 % and sending me the copies of relevant references. 
%%I am grateful to M. Pietrzyk for her help with the 
%%bibliographical search at the initial stage of this work. 
I  also thank NORDITA (Copenhagen) 
for kind hospitality in  August of 1998  
when a part of this work was done. 
It is my pleasure to thank the Organizers of the XXXI Symposium 
on Mathematical Physics in Toru\'n (Poland) for their warm  
hospitality and the financial support.

%and the Institute of Theoretical Physics of the F. Schiller 
%University in Jena for offering me excellent working conditions 
%during 

}

%%%%%%%%%%% Insert your bibliography below %%%%%%%%%%

\end{document}